\documentclass[12pt]{article}

\usepackage{epsfig}
\usepackage{amsbsy}
\usepackage{amsmath}

\begin{document}

\title{Physics from HBT radii}
\author{K.Zalewski
\\ M.Smoluchowski Institute of Physics
\\ Jagellonian University, Cracow\footnote{Address: Reymonta 4, 30 059 Krakow,
Poland, e-mail: zalewski@th.if.uj.edu.pl. This work has been partly supported
by the Polish Ministry of Education and Science grant 1P03B 045 29(2005-2008).
}
\\ and\\ Institute of Nuclear Physics, Cracow}
\maketitle

\begin{abstract}
An approximate formula connecting the true and the HBT homogeneity regions in
multparticle production processes is derived. It implies that when calculating
the HBT radii one should use the center of mass systems of the pairs rather
than the now popular LCMS system. A discussion of several simple examples
clarifies the potential and limitations of the HBT method. The even cumulants
of the $\textbf{X}$-distribution, including the HBT radii, can be determined
for each homogeneity region, but the relative positions of the homogeneity
regions are unconstrained. This makes the HBT radii of little use for
calculating quantities dependent on the interparticle interactions in
coordinate space.
\end{abstract}
\noindent PACS numbers 25.75.Gz, 13.65.+i \\Bose-Einstein correlations,
interaction region determination. \vspace{0.5in}

Almost fifty years ago, in a famous paper known as GGLP \cite{GGL}, a method of
using momentum distributions for pairs of identical pions to estimate the sizes
of the interaction regions, i.e. of the regions where the hadrons are produced
in multiparticle production processes, has been described. Such radii,
determined from momentum distributions of identical particles, have been later
called, not very appropriately \cite{KOP}, HBT radii. GGLP assumed that hadron
production happens instantly and simultaneously at some time $t=0$ and that
there are no correlations between the momenta of the hadrons and their
production points. Considering simultaneously the momentum of a particle and
its production point implies a quasiclassical approximation, but it can be made
plausible \cite{ZAL1} that this approximation is good for heavy ion scattering
and acceptable also for the other multiparticle production processes. Let us
denote by $p_1$ and $p_2$ the four-momenta of the two identical pions in the
pair\footnote{We discuss pions for definiteness, but the discussion applies to
any spin zero bosons.}, by $x_1$ and $x_2$ the space-time positions of their
production points and introduce the notation

\begin{equation}\label{}
  K = \frac{1}{2}(p_1+p_2);\quad q = p_1 - p_2;\quad X = \frac{1}{2}(x_1 +
  x_2);\quad y = x_1 - x_2.
\end{equation}
Note that $p_1^2 = p_2^2 = m^2$ implies

\begin{equation}\label{onmass}
  K_0 =
  \frac{1}{\sqrt 2}\sqrt{E_k^2 + \frac{1}{4}\textbf{q}^2 + \sqrt{(E_k^2 + \frac{1}{4}\textbf{q}^2)^2  -
  (\textbf{K}\cdot \textbf{q})^2}};\quad q_0 = {\mbox{\boldmath$\beta $}}\cdot \textbf{q},
\end{equation}
where $E_k = \sqrt{\textbf{K}^2 + m^2}$ equals $K_0$ at $\textbf{q}=0$, and
${\mbox{\boldmath$\beta $}} = \frac{\textbf{K}}{K_0}$ is the velocity of the
pair. GGLP found $R_{HBT}$ from the distribution of $q^2$.

Some twenty five year later Pratt \cite{PRA1} described a model with an
exploding source. In this model, the momentum distribution of particles depends
on the production point. Due to this correlation, the HBT radius differs
significantly from the true radius, which is known from the input. For fixed
$\textbf{K}$ the result was that, while the true radius $R$ does not depend on
$|\textbf{K}|$, the HBT radius decreases from $\sqrt{\frac{2}{3}}R$ for
$\textbf{K}=0$ to zero when $|\textbf{K}|$ tends to infinity. This finding was
generalized by Bowler \cite{BOW}, who pointed out that in general, whenever
there are strong correlations between momenta and production points, $R_{HBT}$
measures only the region where particles of similar momentum are produced.
Sinyukov \cite{SIN}, \cite{AKS} associated these regions with the homogeneity
regions considered in hydrodynamics and the name homogeneity region got
generally accepted.

Most models in their simplest form, i.e. without final state interactions,
resonance decays etc., find the $n$-particle momentum distributions for
identical pions from formulae equivalent to \cite{KAR}

\begin{equation}\label{karczm}
  P(\textbf{p}_1,\ldots,\textbf{p}_n) = C_n \sum_Q \prod_{j=1}^n \rho(\textbf{p}_j;\textbf{p}_{Qj}),
\end{equation}
where the summation is over all the permutations $j \rightarrow Qj$ of the
indices $j=1,\ldots,n$, $C_n$ are normalization constants and
$\rho(\textbf{p}_1;\textbf{p}_2)$ is some time independent, single particle
density matrix. Thus $\rho(\textbf{p}_1;\textbf{p}_2)$ determines all the
momentum distributions. The best \cite{HCS} quantum mechanical analogue of the
classical phase space density is the Wigner function
$W(\textbf{X},\textbf{K})$, which is a Fourier transform of
$\rho(\textbf{p}_1;\textbf{p}_2)$. Following most models, we assume further
that, $W(\textbf{X},\textbf{K})$ defines the HBT $\textbf{X}$-distribution for
the homogeneity region corresponding to a given $\textbf{K}$.

Measuring all the momentum distributions is not enough to determine
$\rho(\textbf{p}_1;\textbf{p}_2)$. As easily seen from (\ref{karczm}), the
observable momentum distributions do not change \cite{BIZ}, \cite{ZAL2}, under
the transformation\footnote{The full group of transformations includes also
$\rho(\textbf{p}_1;\textbf{p}_2)\rightarrow \rho^*(\textbf{p}_1;\textbf{p}_2)$,
but this corresponds to the space inversion of the interaction region and is of
no interest in the present context.}

\begin{equation}\label{rhopri}
  \rho(\textbf{p}_1;\textbf{p}_2) \rightarrow \rho_f(\textbf{p}_1;\textbf{p}_2) = e^{i\left[f(p_1)-
  f(p_2)\right]}\rho(\textbf{p}_1;\textbf{p}_2),
\end{equation}
where $f(p)$ is an arbitrary real-valued function of the four-vector $p$, and
$p_0 = \sqrt{\textbf{p}^2+m^2}$. This transformation does not affect the HBT
radii\footnote{More generally, all the even cumulants of
$W(\textbf{X},\textbf{K})$, at $\textbf{K}$ fixed, remain unchanged. The proof
is a simple extension of the proof given for the HBT radii in \cite{ZAL2}.} of
the homogeneity regions, but it can shift and deform these regions.

An important concept introduced in \cite{PRA1} is the emission function
$S(X,K)$. Assuming chaotic sources, i.e. no interference between particles
produced at different moments of time, this is related to the density matrix in
the momentum representation by the formula\footnote{Sometimes formulae
equivalent to (\ref{rhoemi}) with $\rho(\textbf{p}_1;\textbf{p}_2)$ replaced by
$\rho^*(\textbf{p}_1;\textbf{p}_2)$ are used. The advantage of the present
convention is that for $S(X,K) = \delta (t)g(\textbf{X},\textbf{K})$ function
$g$ is just the Wigner function.}

\begin{equation}\label{rhoemi}
  \rho(\textbf{p}_1;\textbf{p}_2) = \int\!\!d^4X\;S(X,K)e^{iqX}.
\end{equation}
In the spirit of the quasiclassical approximation, using the well tested (cf.
e.g. \cite{WIH} and references given there) mass shell approximation

\begin{equation}\label{masapr}
  K_0 = E_k \equiv \sqrt{m^2 + \textbf{K}^2},
\end{equation}
one can interpret $S(X,K)$ as the time-dependent distribution of the pairs of
vectors $\{\textbf{X},\textbf{K}\}$. The time independent homogeneity region
for each $\textbf{K}$ can be obtained by integrating over time

\begin{equation}\label{rhotru}
  p(\textbf{X}|\textbf{K}) = \int\!\!dt\;S(X,K),
\end{equation}
where the notation stresses that we are interested in the distribution of
$\textbf{X}$ for given $\textbf{K}$.  At this point the mass shell
approximation (\ref{masapr}) is necessary in order to make $K_0$, and
consequently $S(X;K)$ and $p(\textbf{X}|\textbf{K})$, independent of
$\textbf{q}$.

Further we will call $p(\textbf{X}|\textbf{K})$ true distribution in order to
distinguish it from the HBT distribution given by the Wigner function.
Invariance (\ref{rhopri}) means that for a given set of momentum distributions
there is a variety of HBT $\textbf{X}$-distributions which all correspond to
the same fit to the data. One way of choosing among them is to specify the
emission function.

It was soon noticed \cite{PRA2} that for a given
$\rho(\textbf{p}_1;\textbf{p}_2)$ there is an infinity of solutions for $S$ and
that, in particular, a small long-lived source may be undistinguishable from a
large short-lived source. Thus, the information about interaction regions
obtained from momentum measurements is rather incomplete.

The standard relation between the density matrix and the Wigner function yields
from (\ref{rhoemi})

\begin{equation}\label{}
  W(\textbf{X},\textbf{K}) =
  \int\!\!\frac{d^3q}{(2\pi)^3}\int\!\!d^4X'\;S(X',K)e^{i\textbf{q}\cdot
  (\textbf{X}-\textbf{X}'+{\mbox{\boldmath$\beta $}} t)},
\end{equation}
where $q_0$ has been eliminated using (\ref{onmass}). We invoke now the mass
shell approximation (\ref{masapr}), which makes $S$ independent of
$\textbf{q}$. Thus, the integration $d^3q$ gives
$(2\pi)^3\delta(\textbf{X}-\textbf{X}'+{\mbox{\boldmath$\beta $}} t)$ and the
integration $d^3X'$ can be performed. The result is

\begin{equation}\label{rhohbt}
  W(\textbf{X},\textbf{K}) = \int\!\!dt\;S(\textbf{X}+{\mbox{\boldmath$\beta $}} t,t,K).
\end{equation}
The comparison of the true interaction region with the HBT one reduces to the
comparison of the integrals in (\ref{rhotru}) and (\ref{rhohbt}).

The integrals in (\ref{rhotru}) and (\ref{rhohbt}), and consequently the true
and HBT homogeneity regions, coincide only\footnote{Except for the trivial case
when $S(X,K) \sim \delta(t)$.} for ${\mbox{\boldmath$\beta $}} = \textbf{0}$.
This can be achieved by working with pairs which have the same velocity
${\mbox{\boldmath$\beta $}}$ and using the reference frame where
${\mbox{\boldmath$\beta $}} = \textbf{0}$.

We summarize our findings:
\begin{itemize}
  \item The HBT results are credible only for single homogeneity regions, i.e. at given $\textbf{K}$.
  \item For each homogeneity region, only the pairs with the same
velocity, i.e. with the same $\textbf{K}$, are considered and one should use
the reference frame, where this velocity is zero. There, the true and the HBT
homogeneity regions coincide which is not the case for other frames, as for
instance for the now popular (cf e.g. the review \cite{LPS}) LCMS frame. An
additional pragmatic argument in favor of the rest frame is that the effects of
the final state interactions are the simplest there (cf. e.g. \cite{DAP} and
the references given there).
\item The true radii of the homogeneity regions, more generally all the even cumulants of
their $\textbf{X}$-distributions, can be obtained from the HBT analysis of the
momentum distributions. For instance, when $p(\textbf{X}|\textbf{K}) =
p(-\textbf{X}|\textbf{K})$ all the odd cumulants vanish and, therefore,
$p(\textbf{X}|\textbf{K})$ can be measured by the HBT method.
\item The space distribution of the centers of the homogeneity regions $\langle
\textbf{X}\rangle(\textbf{K})$ is unconstrained by the HBT analysis, but once
it is fixed, all the functions $p(\textbf{X}|\textbf{K})$ can be determined
\cite{BIZ}, \cite{ZAL2}.
\end{itemize}

Let us discuss some simple examples illustrating these features of the HBT
method. In order to avoid unessential complications, we will consider one space
dimension and very simple emission functions constructed from
$\delta$-functions and step functions. They violate the Heisenberg uncertainty
relations, but all these calculations can be repeated in three dimensions using
Gaussians and the results are qualitatively the same.

Let us first consider the emission function

\begin{equation}\label{}
  S_1(X,K;t_0,a,b,\kappa) = \delta(t - t_0)\Theta_{a,b}(\textbf{X})\delta(\textbf{K} - \kappa),
\end{equation}
where $t_0$, $a < b$ and $\kappa$ are real constants. Function $\Theta_{a,b}(x)
= \frac{1}{b-a}$ for $a < x < b$ and zero outside this interval. For this
emission function

\begin{eqnarray}\label{rhomod}
\rho(\textbf{p}_1;\textbf{p}_2) &=&
\delta(\textbf{K}-\kappa)e^{-i\textbf{q}(\frac{a+b}{2} - {\mbox{\boldmath$\beta
$}} t_0)}
\frac{2\sin |\textbf{q}|\frac{b-a}{2}}{|\textbf{q}|(b-a)}\\
W(\textbf{X},\textbf{K}) &=& \Theta_{a,b}(\textbf{X}+ {\mbox{\boldmath$\beta
$}} t_0)\delta(\textbf{K}-\kappa).
\end{eqnarray}
It is seen that the segment $a < \textbf{X} < b$ got shifted to
$a-{\mbox{\boldmath$\beta $}} t_0 < \textbf{X} < b - {\mbox{\boldmath$\beta $}}
t_0$. Acording to (\ref{rhomod}) this shift is due to the phase factor in the
density matrix. Since in the formula for the two-particle correlation function
only the absolute value of $\rho$ appears, experiment is blind to such shifts.
This remains true also when many-particle correlation functions are measured,
even when all the measurements are performed with perfect precision \cite{BIZ},
\cite{ZAL2}. The fact that the position of the center of the interaction region
cannot be found from the HBT analysis of the momentum distributions is, of
course, well known (see e.g. \cite{WIH}), but one should also keep in mind that
when the emission function is given the density matrix, the Wigner function
and, consequently, all the HBT homogeneity regions are unambiguously defined.

As the next example we take

\begin{equation}\label{}
  S_2(X,K) = \frac{1}{2}\left[S_1(X,K;t_0,-a,0,\kappa_1) +
  S_1(X,K;t_0,0,a,\kappa_2)\right]
\end{equation}
Using the previous example, it is seen that the segment $-a < \textbf{X} < a$
is mapped onto two segments: $-a-{\mbox{\boldmath$\beta $}}_1t_0 < \textbf{X} <
-{\mbox{\boldmath$\beta $}}_1 t_0$ and $-{\mbox{\boldmath$\beta $}}_2t_0 <
\textbf{X} < a-{\mbox{\boldmath$\beta $}}_2t_0$. Let us choose

\begin{equation}\label{trainv}
  f(p) = -\textbf{b}\cdot \textbf{p} - \frac{1}{2}c\textbf{p}^2,
\end{equation}
where $\textbf{b}$ is an arbitrary vector and $c$ an arbitrary constant, and
make the transformation (\ref{rhopri}). Each HBT homogeneity region gets
shifted by $\textbf{b} + c\textbf{K}$. For $\kappa_1~\neq ~\kappa_2$, by a
suitable choice of $c$, one can obtain any prescribed distance between the
centers of the two segments. The true length of the interaction region, as seen
from $S$, is $2a$. The length of the transformed interaction region, which
follows just as well from the data on momentum distributions, can be any number
not smaller than $a$. What is the way out?  One has to invoke the homogeneity
regions. For each $\textbf{K}$ separately, the length of the segment where
particles with this value of $\textbf{K}$ are produced, i.e. of the homogeneity
region, is reproduced correctly. The positioning of the homogeneity regions
corresponding to different values of $\textbf{K}$, however, is beyond control
when the momentum distributions are the only input. In three dimensions it is
easy to prove \cite{BIZ}, \cite{ZAL2} that by a suitable choice of function
$f(p)$ in (\ref{rhopri}), the positions of the centers of the homogeneity
regions $\langle \textbf{X} \rangle(\textbf{K})$ can be changed into $\langle
\textbf{X} \rangle(\textbf{K}) + \textbf{g}(\textbf{K})$, where $\textbf{g}$ is
an arbitrary differentiable function of $\textbf{K}$.

As an amusing example in three dimensions let us consider the models where
$K_\mu \approx \lambda X_\mu$, and $\lambda$ is a constant \cite{CSO},
\cite{BIZ2}. They correspond to emission functions

\begin{equation}\label{}
  S(X,K) = \delta^3(\textbf{X}- {\mbox{\boldmath$\beta $}} t)\overline{S}(\textbf{X},t,K),
\end{equation}
where $\overline{S}$ is some function which does not affect the singularity
introduced by the $\delta^3$. The corresponding Wigner function is

\begin{equation}\label{}
  W(\textbf{X},\textbf{K}) = \delta^3(\textbf{X})\int\!\!dt\;\overline{S}({\mbox{\boldmath$\beta $}} t,t,K).
\end{equation}
Thus, the HBT interaction region reduces to one point. To be sure: such models,
when properly used, are quite successful, but finding the interaction region
from the density matrix in the momentum representation is their misuse.

As our last model consider the emission function

\begin{equation}\label{}
  S(X,K) = \Theta_{0,a}(t)\delta(\textbf{X})\delta(\textbf{K}-\kappa).
\end{equation}
This corresponds to the Wigner function
\begin{equation}\label{}
  W(\textbf{X},\textbf{K}) = \frac{1}{{\mbox{\boldmath$\beta $}}}\Theta_{0,a}(-\textbf{X}/{\mbox{\boldmath$\beta $}})
  \delta(\textbf{K} - \kappa) =
  \Theta_{-{\mbox{\boldmath$\beta $}a,0} }(\textbf{X})\delta(\textbf{K} - \kappa).
\end{equation}
The length of the true interaction region is zero, while the length of the HBT
interaction region is ${\mbox{\boldmath$\beta $}} a$. Thus, in order to get the
true length from the HBT analysis one must use the reference frame where
$\textbf{K} = \textbf{0}$. We conclude that, working with all the pairs which
have a given velocity, one should measure their homogeneity region in their
rest frame. Let us consider some further implications of our analysis:

If there are no position-momentum correlations, all the homogeneity regions,
measured in the respective rest frames, are equivalent to each other and to the
overall interaction region. The ambiguity (\ref{rhopri}) reduces to a lack of
information about the position of the center of the interaction region
\cite{ZAL2}. Thus, in this case the HBT method works very well. One should keep
in mind however that, as illustrated by our second model, the independence of
the homogeneity region on $\textbf{K}$ is a necessary, but not a sufficient
condition for the absence of position-momentum correlations.

If there are no interparticle interactions in coordinate space, the HBT
homogeneity regions may be all we need. E.g. the entropy of a gas of
noninteracting particles (in the quasiclassical approximation) depends on the
accessible phase space volume. This can be calculated by integrating over
coordinate space at fixed momentum and then integrating over momenta. The first
integration gives just the HBT volume of the homogeneity region, though one
should keep in mind the ambiguity (\ref{rhopri}). If there are interparticle
interactions in coordinate space, however, it may make a lot of difference
whether the homogeneity regions are on top of each other, or scattered over
space. Thus, the use of the HBT radii to calculate the entropy of a gas of
interacting particles, or of their mean free paths, is risky.

For ${\mbox{\boldmath$\beta $}} = \textbf{0}$ the true space density
$\int\!\!dt\;S(X,K)$ can be unambiguously obtained from
$\rho(\textbf{p}_1,\textbf{p}_2)$ by inverse Fourier transformation. Thus, all
the ambiguity in the determination of the homogeneity regions, including their
absolute positions, results from (\ref{rhopri}) and disappears when $f(p)$ is
fixed. Conversely, when the true space density is known, it yields
unambiguously $\rho(\textbf{p}_1;\textbf{p}_2)$ and thus it fixes $f(p)$.

The ambiguity in the determination of the space-time density $S$ follows from
the irreversibility of transformation (\ref{rhoemi}) and persists even when
$f(p)$ is fixed, i.e. when $\rho(\textbf{p}_1;\textbf{p}_2)$ is known.

\Large{\textbf{Acknowledgements}}\normalsize

The author thanks A. Bialas, K.Fia{\l}kowski and W. Florkowski for helpful
comments.

\end{document}